\documentclass[a4paper,12pt]{article}
\usepackage[pctex32]{graphics}
\usepackage{amssymb,amsmath}
\usepackage{amsmath,amssymb}
\usepackage{latexsym}
\usepackage{epsfig}
\usepackage[english]{babel}

\begin{document}

\topmargin 0pt \oddsidemargin 0mm

\newcommand{\be}{\begin{equation}}
\newcommand{\ee}{\end{equation}}
\newcommand{\ba}{\begin{eqnarray}}
\newcommand{\ea}{\end{eqnarray}}

\begin{titlepage}

\vspace{5mm}
\begin{center}
{\Large \bf Quasinormal modes around the BTZ black hole \\
at the tricritical generalized massive gravity}

\vskip .6cm

\centerline{\large
 Yong-Wan Kim $^{1,a}$,  Yun Soo Myung$^{1,b}$,
and Young-Jai Park$^{2,3,c}$}

\vskip .6cm

{$^{1}$Institute of Basic Science and School of Computer Aided
Science, \\Inje University, Gimhae 621-749, Korea \\}

{$^{2}$Department of Physics and Center for Quantum Spacetime, \\
Sogang University, Seoul 121-742, Korea}\\

{$^{3}$Department of Global Service Management,\\
Sogang University, Seoul 121-742, Korea}

\end{center}

\begin{center}

\underline{Abstract}
\end{center}
Employing the operator method, we obtain  log-square  quasinormal
modes and frequencies of a graviton  around the BTZ black hole  at
the tricritical point of the generalized massive gravity. The
log-square quasinormal frequencies are  also obtained  by
considering a  finite temperature conformal field theory. This shows
the AdS/LCFT correspondence at the tricritical point approximately.
We discuss a truncation process to the unitary theory on the BTZ
black hole background.

\vskip .6cm

\noindent
PACS numbers: 04.70.Bw, 04.30.Nk, 04.60.Kz, 04.60.Rt \\

\noindent Keywords: tricritical gravity; logarithmic quasinormal
modes

\vskip 0.8cm

\vspace{15pt} \baselineskip=18pt

\noindent $^a$ywkim65@gmail.com\\
\noindent $^b$ysmyung@inje.ac.kr \\
\noindent $^c$yjpark@sogang.ac.kr

\thispagestyle{empty}
\end{titlepage}

\newpage
\section{Introduction}
Critical gravities have  been the subject of active interest because
they were considered as toy models for quantum
gravity~\cite{Li:2008dq,Lu:2011zk,Deser:2011xc,Porrati:2011ku}.
According to the AdS/LCFT correspondence, one finds that a rank-2
logarithmic conformal field theory (LCFT)  is dual to a critical
gravity~\cite{Grumiller:2008qz,Myung:2008dm,Maloney:2009ck}.
However, one has to deal with the non-unitarity issue of these
theories because the LCFT is in general non-unitary.

Recently, a higher-derivative critical (polycritical) gravity was
introduced to provide multiple critical points~\cite{Nutma:2012ss}
which might be described by  a higher-rank LCFT.  The rank of the
LCFT refers to the dimensionality of the Jordan cell. The LCFT dual
to critical gravity has rank-2 and thus, an operator has one
logarithmic partner. The LCFT dual to tricritical gravity has rank-3
and thus, an operator has two logarithmic partners. An odd-rank LCFT
allows for a truncation to a unitary conformal field theory
(CFT)~\cite{Bergshoeff:2012sc}. A six-derivative gravity in three
dimensions was treated as dual to a rank-3
LCFT~\cite{Bergshoeff:2012ev}, while four-derivative critical
gravity in four dimensions was considered as dual to a rank-3
LCFT~\cite{Johansson:2012fs}. Furthermore, it is shown that a
consistent unitary truncation of polycritical gravity was possible
to occur at the linearized level for odd
rank~\cite{Kleinschmidt:2012rs}.   On the other hand, a non-linear
critical gravity of rank-3 in three and four dimensions was
investigated in~\cite{Apolo:2012vv}, showing that truncation which
appears to be unitary at the linear level might be inconsistent at
the non-linear level. It worth noting that a tricritical gravity was
first mentioned in the six-derivative gravity in six
dimensions~\cite{LPP}.

It is important to note that one can construct a rank-3 parity odd
theory in the simple context of four-derivative gravity as
generalized massive gravity (GMG) in the three dimensional anti-de
Sitter (AdS$_3$) spacetimes~\cite{LS}. In fact, the GMG is a combination of
topologically massive gravity (TMG)~\cite{Deser:1981wh} and new
massive gravity (NMG)~\cite{Bergshoeff:2009hq}. There exists one
parity-odd tricritical point in the GMG parameter space where the
theory propagates one left-moving massless graviton as well as
right-moving massless graviton, and two logarithmic modes
associated with left-movers known as  log and log$^2$ boundary
behaviors on the AdS$_3$ background. Its dual theory is a rank-3
LCFT~\cite{Grumiller:2010tj}. A truncation of tricritical GMG  is
made by imposing $Q_L=0$ with $Q_L$ the Abbott-Deser-Tekin
charge~\cite{Bergshoeff:2012ev}. After truncation, one has found
the left-moving sector with a CFT which is unitary, in addition to
the right-moving massless sector which dictates the chiral
gravity.

Before we proceed, we would like to mention the following difference
in AdS/LCFT correspondence:

\indent $\bullet$ tricritical gravity on the AdS$_3 \rightarrow$
 a rank-3 zero temperature LCFT

\indent $\bullet$ tricritical gravity on the BTZ black hole $
\rightarrow$ a rank-3 finite temperature~LCFT.

In this work, to confirm the AdS/LCFT
correspondence~\cite{Birmingham:2001pj}, we obtain
log$^2$-quasinormal modes and frequencies of a graviton around the
BTZ black hole (instead of the AdS$_3$) at the tricritical GMG by
employing the operator method. In order to obtain the quasinormal
modes precisely, one needs to know a rank-3 finite temperature LCFT
while a rank-2 finite temperature LCFT was known in Ref.
\cite{Myung:1999nd}. Here, we show that the log$^2$-quasinormal
frequencies are also obtained by considering a finite temperature
CFT~\cite{Sachs}.

\section{Generalized massive gravity}

We consider the generalized massive gravity (GMG) action
 \be\label{gmg}
 S_{\rm GMG}=\frac{1}{16\pi G}\int d^3x~\sqrt{-g}\left[\sigma R-2\lambda +\frac{1}{m^2}K+\frac{1}{\mu}L_{\rm CS}\right],
 \ee
where $K$ ($L_{\rm CS}$) is the new massive gravity (NMG) term (the
Chern-Simons term) given by
 \ba
 K&=&R_{\mu\nu}R^{\mu\nu}-\frac{3}{8}R^2,\\
 L_{\rm CS}&=&\frac{1}{2}\epsilon^{\mu\nu\rho}\Gamma^\alpha_{\mu\beta}
  \left[\partial_\nu\Gamma^\beta_{\alpha\rho}+\frac{2}{3}\Gamma^\beta_{\nu\gamma}\Gamma^\gamma_{\rho\alpha}\right].
 \ea
Here $m$ and  $\mu$ are the two mass parameters, while $\sigma$ is a
dimensionless sign parameter which takes $+1$ for our purpose. We also use the convention of $\epsilon^{\rho uv}=1/\sqrt{-g}$~\cite{Grumiller:2009mw}.
Replacing $m^2$ by $-m^2$ leads to the action in~\cite{LS}.  Also,
$\lambda$ is the cosmological constant. Its equation of motion takes
the form \be G_{\mu\nu}+\lambda
g_{\mu\nu}+\frac{1}{2m^2}K_{\mu\nu}+\frac{1}{\mu}C_{\mu\nu}=0, \ee
where
 \ba
 G_{\mu\nu}&=&R_{\mu\nu}-\frac{1}{2}g_{\mu\nu}R, \\
 K_{\mu\nu}&=&-\frac{1}{2} \nabla^2R g_{\mu\nu}-\frac{1}{2}
 \nabla_\mu\nabla_\nu R +2 \nabla^2 R_{\mu\nu} \nonumber  \\
 &+&4R_{\mu\alpha\nu \beta}R^{\alpha \beta} -\frac{3}{2}
 RR_{\mu\nu}-R_{\alpha \beta}R^{\alpha \beta}
 g_{\mu\nu}+\frac{3}{8}R^2 g_{\mu\nu},
 \ea
and the Cotton tensor is given by
 \be
 C_{\mu\nu}=\epsilon_{\mu}^{~\alpha\beta}\nabla_{\alpha}
 \Big(R_{\beta\nu}-\frac{1}{4} g_{\beta\nu} R\Big).
 \ee In this work,
we consider the BTZ black hole in the Schwarzschild coordinates
\be \label{sle} ds^2_{\rm BTZ}=-\Big(-{\cal
M}+\frac{r^2}{\ell^2}\Big)dt^2+\frac{dr^2}{-{\cal
M}+\frac{r^2}{\ell^2}}+r^2d\phi^2 \ee whose horizon is located at
$r_+=\ell \sqrt{{\cal M}}$.  The ${\cal M}=-1$ case corresponds to
the AdS$_3$ spacetimes, while ${\cal M}=1$ provides a unity mass
of the BTZ black hole. In these cases, one finds a relation among
$m^2$, $\lambda$, and $\Lambda$ as
 \be
 m^2=\frac{\Lambda^2}{4(\lambda-\Lambda)},~~ \Lambda=-1
 \ee with
$\ell=1$. Also, the left-temperature, and right-temperature, and
Hawking temperature are the same as \be
T_L=T_R=T_H=\frac{1}{2\pi}. \ee The line element (\ref{sle}) is
expressed in terms of global coordinates
 \ba\label{btz}
  ds^2
  &=&-\sinh^2(\rho)d\tau^2+\cosh^2(\rho)d\phi^2+d\rho^2,
 \ea
where we have introduced the radial coordinate $r=\cosh\rho$ such
that the event horizon of $r=r_+=1$ is located at $\rho=0$, while
the infinity is at $\rho=\infty$.   Introducing the light cone
coordinates $u/v=\tau\pm \phi$, the line element  becomes
 \ba\label{lc}
 ds^2&=&\bar{g}_{\mu\nu}dx^\mu dx^\nu \nonumber \\
     &=& \frac{1}{4}du^2-\frac{1}{2}\cosh(2\rho)dudv+\frac{1}{4}dv^2+d\rho^2.
 \ea
Then, the metric tensor (\ref{lc}) admits the Killing vector fields
$L_k$, $k=0,-1,1$ for local SL$(2,R)_{\rm L}\times$SL$(2,R)_{\rm R}$ algebra as
 \be
 L_0=-\partial_u,~~L_{-1/1}=e^{\mp u}
 \Big[-\frac{\cosh(2\rho)}{\sinh(2\rho)}\partial_u
 -\frac{1}{\sinh(2\rho)}\partial_v \mp \frac{1}{2} \partial_\rho\Big],
 \ee
and $\bar{L}_0,\bar{L}_{-1/1}$ are similarly obtained by
interchanging $u$ and $v~ (u \leftrightarrow v)$. Locally, they
form a basis of the Lie algebra SL$(2,R)$ as
 \be
 [L_0,L_{\pm 1}]=\mp L_{\pm 1},~~[L_1,L_{-1}]=2L_0,
 \ee
which will be used to generate the whole tower of quasinormal modes
in three dimensions.

Now, we are going to  expand
$g_{\mu\nu}=\bar{g}_{\mu\nu}+h_{\mu\nu}$ around the BTZ background
in Eq. (\ref{lc}) and choose the transverse-traceless (TT) gauge
 \be \label{TT}
 \bar{\nabla}_{\mu}h^{\mu\nu}=0,~~h^{\mu}_{~\mu}=0.
 \ee
Here we wish to mention that the TT gauge is allowed, thanks to
$\delta R(h)=0$  which is obtained by tracing both sides of the
linearized Einstein equation. Under the TT gauge, the linearized
Einstein equation becomes the fourth-order  differential equation
 \be\label{4theom}
 (\bar{\nabla}^2-2\Lambda)
 \Bigg[\bar{\nabla}^2h_{\mu\nu}+\frac{m^2}{\mu}\epsilon^{~\alpha\beta}_\mu\bar{\nabla}_\alpha h_{\beta\nu}
 +\left(m^2-\frac{5}{2}\Lambda\right)h_{\mu\nu}\Bigg]=0.
 \ee
 Introducing four  mutually commuting operators
of
 \ba \label{meq}
 (D^{L/R})^\beta_\mu &=& \delta^\beta_\mu\pm\epsilon_\mu^{~\alpha\beta}\bar{\nabla}_\alpha,\nonumber\\
 ~~(D^{m_i})^\beta_\mu &=& \delta^\beta_\mu
            +\frac{1}{m_i}\epsilon_\mu^{~\alpha\beta}\bar{\nabla}_\alpha,~~~(i=1,2),
 \ea
the linearized equation of motion (\ref{4theom}) can be written to
be compactly
 \be \label{geq}
 \Big(D^RD^LD^{m_1}D^{m_2}h\Big)_{\mu\nu}=0.
 \ee
Here, the mass parameters are given by
 \ba
 m_1&=&\frac{m^2}{2\mu}+\sqrt{\frac{m^4}{4\mu^2}-m^2+\frac{1}{2}},\nonumber\\
 m_2&=&\frac{m^2}{2\mu}-\sqrt{\frac{m^4}{4\mu^2}-m^2+\frac{1}{2}}.
 \ea
The parameter space is shown in~\cite{Bergshoeff:2012ev}. The two
critical lines appear when $m_1=m_2$. The NMG and TMG limits of
the GMG are on the $\frac{1}{m^2}(x)$-axis and
$\frac{1}{\mu}(y)$-axis, respectively. When a critical line
intersects with one of them, either critical TMG or critical NMG
is recovered. We are interested in two tricritical points \ba
{\rm point}~ &1&:m^2=2\mu=\frac{3}{2}, \\
{\rm point}~ &2&:m^2=-2\mu=\frac{3}{2}. \ea At the point 1 of
$m_1=m_2=1$, ${\cal D}^{m_1}$ and ${\cal D}^{m_2}$ degenerate with
${\cal D}^{L}$, while
 at the  point 2 of $m_1=m_2=-1$, ${\cal D}^{m_1}$ and ${\cal D}^{m_2}$
degenerate with ${\cal D}^{R}$. The presence of two tricritical
points is a main feature of the GMG,  but it is not a feature of
the TMG or NMG. This implies that there is no tricritical point in
the context of the TMG or NMG.  Hereafter, we will focus on the
tricritical point 1 because results for the point 2 could be
obtained by exchanging $L$ and $R$.

\section{Log-square quasinormal modes at the tricritical GMG}

We start with a first-order massive equation
 \ba \label{eql1}
 (D^{M}h)_{\mu\nu} = 0 \to
 \epsilon_\mu^{~\alpha\beta}\bar{\nabla}_\alpha
 h_{\beta\nu}+Mh_{\mu\nu}=0,
 \ea
 where $M=m_{i}$.
This can be solved with the TT gauge as~\cite{SS,MKMP}
 \be \label{hmassive}
 h^M_{\mu\nu}=e^{-ik(\tau+\phi)-2h_L\tau} (\sinh\rho)^{-2h_L}(\tanh\rho)^{-ik}
 \left(
  \begin{array}{ccc}
    1 & 0 & \frac{2}{\sinh(2\rho)} \\
    0 & 0 & 0 \\
    \frac{2}{\sinh(2\rho)} & 0 & \frac{4}{\sinh^2(2\rho)}\\
  \end{array}
 \right),
 \ee
 where $h_L$ is  the conformal weight of graviton with mass $M$  given by
 \be \label{conw}
 h_L=\frac{M-1}{2},~~M \ge 1.
 \ee
As was pointed out in~\cite{SS}, the highest weight condition of
$L_1 h^M_{\mu\nu}=\bar{L}_1 h^M_{\mu\nu}=0$ which is suited to
reproduce the normalizable modes in AdS$_3$ spacetimes is too
strong in the BTZ black hole background because the descendants of
such highest weight modes have imaginary $\phi$-momentum. This
problem could be resolved when imposing a weaker condition
so-called the chiral highest weight condition of $\bar{L}_1
h^M_{\mu\nu}=0$ which is compatible with the TT gauge condition
(\ref{TT}). This allows for real $\phi$-momentum.

For $M=1(=m_1=m_2)$, the solution is reduced to
 \be \label{hmassless}
 h^{M=1}_{\mu\nu}=h^L_{\mu\nu}=e^{-ik(\tau+\phi)}(\tanh\rho)^{-ik}
 \left(
  \begin{array}{ccc}
    1 & 0 & \frac{2}{\sinh(2\rho)} \\
    0 & 0 & 0 \\
    \frac{2}{\sinh(2\rho)} & 0 & \frac{4}{\sinh^2(2\rho)}\\
  \end{array}
 \right),
 \ee
which corresponds to the left-moving solution with $h_L=0$
propagating on the BTZ black hole
background~\cite{MKMP,Myung:2012sh,Kim:2012pt}. This left-moving
solution with the zero conformal weight is a cornerstone to
construct the log-solution at the critical point and the
log$^2$-solution at the tricritical point.

One can now construct the log-solution~\cite{Grumiller:2008qz}  as
 \be\label{logeom1}
 h^{L,{\rm log}}_{\mu\nu} = \partial_{M}h^{M}_{\mu\nu}|_{M=1}=y(\tau,\rho)h^{L}_{\mu\nu},
 \ee
which is responsible for describing the critical GMG. Here
$y(\tau,\rho)$ is defined by
 \be
 y(\tau,\rho)=-\tau-\ln[\sinh(\rho)].
 \ee
The classical stability for this log-solution in the critical NMG
has been explicitly studied in Ref.~\cite{MKMP}. On the other hand,
we need to introduce the log$^2$-solution
 \be\label{loglogeom1}
 h^{L,{\rm log^2}}_{\mu\nu} = \frac{1}{2} \partial^2_{M}h^{M}_{\mu\nu}|_{M=1}=\frac{1}{2}
 y^2(\tau,\rho)h^{L}_{\mu\nu}
 \ee
for describing the tricritical GMG  whose linearized equation is
given by
 \be
 \Big(D^RD^LD^{L}D^{L}h^{L,\log^2}\Big)_{\mu\nu}=0.
 \ee
One can easily check that
 \ba
 && (D^L h^{L,{\rm log^2}})_{\mu\nu} = - h^{L,{\rm log}}_{\mu\nu},\nonumber\\
 && (D^L D^L h^{L,{\rm log^2}})_{\mu\nu}  = h^{L}_{\mu\nu},\nonumber\\
 && (D^L D^L D^L h^{L,{\rm log^2}})_{\mu\nu} =  (D^L h^{L})_{\mu\nu}=0.
 \ea
One of non-trivial tasks on the BTZ black hole is to derive
quasinormal modes of graviton and their frequencies. Usually, one
needs to solve the second-order differential equation to find
quasinormal modes with the ingoing wave at horizon and Dirichlet
boundary condition at infinity. However, if one makes the
second-order equation from the first-order one, sign ambiguity may
appear in the mass term. Hence, it would be better to use the
operator method  to obtain quasinormal modes. According to the
Sachs's proposal~\cite{Sachs}, the logarithmic quasinormal modes
can be constructed by using the operator method as
 \be
 h^{L(n),{\rm log^2}}_{\mu\nu}(u,v,\rho)=
 \Big(\bar{L}_{-1}L_{-1}\Big)^n  h^{L,{\rm
 log^2}}_{\mu\nu}(u,v,\rho),
 \ee
 which implies that all descendants could be obtained from the
chiral highest weight $h^{L,{\rm
 log^2}}_{\mu\nu}$ satisfying  $\bar{L}_1 [h^{L,{\rm
 log^2}}_{\mu\nu}]=0$ by acting $\bar{L}_{-1}L_{-1}$ on it.
Explicitly, the first descendant of $h^{L,{\rm log^2}}_{\mu\nu}$
is given by
 \ba\label{leftsol1}
 h^{L(1),{\rm log^2}}_{\mu\nu}(u,v,\rho) &=&
 \Big(\bar{L}_{-1}L_{-1}\Big) h^{L,{\rm log^2}}_{\mu\nu}(u,v,\rho)\nonumber\\
 &=&\frac{e^{-2\tau}}{2\sinh^2\!\rho}e^{-ik(\tau+\phi)}(\tanh\!\rho)^{-ik}
  \left(
  \begin{array}{ccc}
    f^{L(1)}_{uu} & f^{L(1)}_{uv} & \frac{f^{L(1)}_{u\rho}}{\sinh(2\rho)}\\
    f^{L(1)}_{uv} & 0 & \frac{f^{L(1)}_{v\rho}}{\sinh(2\rho)} \\
    \frac{f^{L(1)}_{u\rho}}{\sinh(2\rho)} & \frac{f^{L(1)}_{v\rho}}{\sinh(2\rho)} & \frac{f^{L(1)}_{\rho\rho}}{\sinh^2(2\rho)} \\
  \end{array}
  \right)_{\mu\nu}, \nonumber\\
 \ea
whose relevant components are given by
 \ba
 f^{L(1)}_{uu}&=& 1+\cosh(2\rho)+[4+2\cosh(2\rho)+ik(3+\cosh(2\rho))]y(t,\rho)\nonumber\\
              &+& (2-k^2+3ik)y^2(\tau,\rho), \nonumber \\
 f^{L(1)}_{uv} &=& 2y(\tau,\rho)\left[1+(1+\frac{ik}{2})y(t,\rho)\right], \nonumber\\
 f^{L(1)}_{u\rho}&=& 2[1+\cosh(2\rho)+4(1+\cosh(2\rho))y(t,\rho)
                       +(2+2\cosh(2\rho)-k^2)y^2(\tau,\rho) \nonumber\\
                 &+& ik(3+\cosh(2\rho))y(t,\rho)(1+y(\tau,\rho))], \nonumber \\
 f^{L(1)}_{v\rho}&=& 4y(\tau,\rho)\left[1+(1+\frac{ik}{2})y(t,\rho)\right], \nonumber\\
 f^{L(1)}_{\rho\rho}&=&  4[1+\cosh(2\rho)+2(2+3\cosh(2\rho))y(t,\rho)
                       +(2+4\cosh(2\rho)-k^2)y^2(\tau,\rho) \nonumber\\
                 &+& ik(3+\cosh(2\rho))y(t,\rho)(1+y(\tau,\rho))].
 \ea

Next, the second descendant of $h^{L,{\rm
log^2}}_{\mu\nu}(u,v,\rho)$ is derived from the operation as
 \ba\label{leftsol2}
 h^{L(2),{\rm log^2}}_{\mu\nu}(u,v,\rho) &=& \Big(\bar{L}_{-1}L_{-1}\Big)^2
                 h^{L,{\rm log^2}}_{\mu\nu}(u,v,\rho) \nonumber\\
&=&\frac{e^{-4\tau}}{2\sinh^4\!\rho}e^{-ik(\tau+\phi)}(\tanh\!\rho)^{-ik}
  \left(
  \begin{array}{ccc}
    f^{L(2)}_{uu} & f^{L(2)}_{uv} & \frac{f^{L(2)}_{u\rho}}{\sinh(2\rho)}\\
    f^{L(2)}_{uv} & f^{L(2)}_{vv} & \frac{f^{L(2)}_{v\rho}}{\sinh(2\rho)} \\
    \frac{f^{L(2)}_{u\rho}}{\sinh(2\rho)} & \frac{f^{L(2)}_{v\rho}}{\sinh(2\rho)} & \frac{f^{L(2)}_{\rho\rho}}{\sinh^2(2\rho)} \\
  \end{array}  \right)_{\mu\nu}, \nonumber\\
 \ea
whose components are explicitly given by
 \ba
 f^{L(2)}_{uu}&=& \frac{1}{4}[145+140\cosh(2\rho)+11\cosh(4\rho)
                  -k^2(27+20\cosh(2\rho)+\cosh(4\rho))]\nonumber\\
                  &+& \frac{1}{4}[274+200\cosh(2\rho)+6\cosh(4\rho)-k^2(163+76\cosh(2\rho)+\cosh(4\rho))]y(\tau,\rho), \nonumber \\
                  &+& (24+12\cosh(2\rho)-k^2(30+7\cosh(2\rho))+k^4)y^2(\tau,\rho)\nonumber\\
                  &+&  \frac{ik}{4}(125+108\cosh(2\rho)+7\cosh(4\rho))\nonumber \\
                  &+&  \frac{ik}{4} [367+220\cosh(2\rho)+5\cosh(4\rho)-8k^2(3+\cosh(2\rho))]y(\tau,\rho) \nonumber \\
                  &+& ik(44+16\cosh(2\rho)-k^2(9+\cosh(2\rho)))y^2(\tau,\rho),
 \ea
 \ba
 f^{L(2)}_{uv}&=& 18+14\cosh(2\rho)+2[26+16\cosh(2\rho)-k^2(5+\cosh(2\rho))]y(t,\rho)\nonumber\\
              &+& 2(12+6\cosh(2\rho)-k^2(7+\cosh(2\rho)))y^2(\tau,\rho)\nonumber\\
              &+& 4ik(2+\cosh(2\rho))+2ik(23+9\cosh(2\rho))y(t,\rho) \nonumber\\
              &+& 2ik(16+5\cosh(2\rho)-k^2)y^2(\tau,\rho),
 \ea
 \ba
 f^{L(2)}_{u\rho}&=&  \frac{1}{2}[177+212\cosh(2\rho)+35\cosh(4\rho)
                  -k^2(27+20\cosh(2\rho)+\cosh(4\rho))]\nonumber\\
                  &+& \frac{1}{2}[358+408\cosh(2\rho)+50\cosh(4\rho)-k^2(167+116\cosh(2\rho)+5\cosh(4\rho))]y(\tau,\rho) \nonumber \\
                  &+& [66+72\cosh(2\rho)+6\cosh(4\rho)-k^2(63+42\cosh(2\rho)+\cosh(4\rho))+2k^4]y^2(\tau,\rho)\nonumber\\
                  &+&  \frac{ik}{2}(133+140\cosh(2\rho)+15\cosh(4\rho))\nonumber \\
                  &+&  \frac{ik}{2} [411+404\cosh(2\rho)+33\cosh(4\rho)-8k^2(3+\cosh(2\rho))]y(\tau,\rho) \nonumber \\
                  &+& ik[103+96\cosh(2\rho)+5\cosh(4\rho)-6k^2(3+\cosh(2\rho))]y^2(\tau,\rho),
 \ea
 \ba
 f^{L(2)}_{vv} &=&
 2[2+10y(\tau,\rho)+(6-k^2)y^2(\tau,\rho)+4iky(\tau,\rho)-5iky^2(\tau,\rho)],
 \ea
 \ba
 f^{L(2)}_{v\rho}&=& 4 \{ 9(1+\cosh(2\rho))+[26+26\cosh(2\rho)-k^2(5+\cosh(2\rho))]y(t,\rho)\nonumber\\
              &+& (12+12\cosh(2\rho)-k^2(7+2\cosh(2\rho)))y^2(\tau,\rho)\nonumber\\
              &+& 2ik(2+\cosh(2\rho))+ik(23+13\cosh(2\rho))y(t,\rho)\nonumber\\
              &+& ik(16+10\cosh(2\rho)-k^2)y^2(\tau,\rho) \},
              \ea
 \ba
 f^{L(2)}_{\rho\rho}&=&  217+284\cosh(2\rho)+67\cosh(4\rho)) -k^2(27+20\cosh(2\rho)+\cosh(4\rho))\nonumber\\
                  &+& [482+616\cosh(2\rho)+134\cosh(4\rho)-3k^2(57+52\cosh(2\rho)+3\cosh(4\rho))]y(\tau,\rho) \nonumber \\
                  &+& 4[48+60\cosh(2\rho)+12\cosh(4\rho)-k^2(34+35\cosh(2\rho)+2\cosh(4\rho))-k^4]y^2(\tau,\rho)\nonumber\\
                  &+&  ik(141+172\cosh(2\rho)+23\cosh(4\rho))\nonumber \\
                  &+&  ik[471+588\cosh(2\rho)+77\cosh(4\rho)-8k^2(3+\cosh(2\rho))]y(\tau,\rho) \nonumber \\
                  &+& 4ik[64+80\cosh(2\rho)+10\cosh(4\rho)-k^2(9+5\cosh(2\rho))]y^2(\tau,\rho).
 \ea
From these expressions, one can deduce the $n^{\rm th}$-order
descendant as
 \ba\label{lqnms}
  h^{L(n),{\rm log^2}}_{\mu\nu}(u,v,\rho)&=&  \Big(\bar{L}_{-1}L_{-1}\Big)^n
                                      h^{L,{\rm log^2}}_{\mu\nu}(u,v,\rho) \nonumber\\
 &=&\frac{e^{-2n\tau}}{2\sinh^{2n}\!\rho}e^{-ik(\tau+\phi)}(\tanh\!\rho)^{-ik}F^{L(n)}_{\mu\nu}(\rho),
 \ea
where $F^{L(n)}_{\mu\nu}(\rho)$ is the corresponding $n^{\rm th}$-
order matrix.  As a result, we read off the log-square quasinormal
frequencies of the graviton  at the tricritical point 1 as
 \be \label{lsqnm}
 \omega^n_L=k-i4\pi T_L n, ~~n\in Z,
 \ee
which is the same expression for the spin-2 graviton $h_{\mu\nu}$ at
the critical point~\cite{Sachs}.

At this stage, it is appropriate to comment on the right-moving
solution and log-square quasinormal modes at the tricritical point
2. The right-moving solution and its corresponding log-square
solution can be easily constructed by the substitution of $u\to v$,
$L\to R~(\phi\rightarrow -\phi,~ k\rightarrow -k$)  in Eqs.
(\ref{hmassive}) and (\ref{hmassless}). Actually, one starts with
 \be \label{imassive}
 h^M_{\mu\nu}=e^{ik(\tau+\phi)-2h_R\tau} (\sinh\rho)^{-2h_R}(\tanh\rho)^{ik}
 \left(
  \begin{array}{ccc}
    0 & 0 & 0 \\
    0 & 1 & \frac{2}{\sinh(2\rho)} \\
    0 & \frac{2}{\sinh(2\rho)} & \frac{4}{\sinh^2(2\rho)}\\
  \end{array}
 \right),
 \ee
 where $h_R$ is  the conformal weight of graviton with mass $M$  given by
 \be \label{iconw}
 h_R=-\frac{M+1}{2},~~ M\le -1.
 \ee
For $M=-1$, $h_{\mu\nu}^M$ leads to $h^R_{\mu\nu}$ \be
\label{imassive}
 h^R_{\mu\nu}=e^{ik(\tau+\phi)} (\tanh\rho)^{ik}
 \left(
  \begin{array}{ccc}
    0 & 0 & 0 \\
    0 & 1 & \frac{2}{\sinh(2\rho)} \\
    0 & \frac{2}{\sinh(2\rho)} & \frac{4}{\sinh^2(2\rho)}\\
  \end{array}
 \right).
 \ee
With the log and log-square solutions of
 \ba
 h^{R,{\rm Log}}_{\mu\nu}&=&-y(\tau,\rho)h^R_{\mu\nu}, \\
  h^{R,{\rm Log2}}_{\mu\nu}&=& \frac{1}{2}y^2(\tau,\rho)h^R_{\mu\nu},
 \ea
the tricritical GMG at the point 2 is described by
 \be
 \Big(D^LD^RD^{R}D^{R}h^{R,\log^2}\Big)_{\mu\nu}=0.
 \ee
One can also easily check that
 \ba
 && (D^R h^{R,{\rm log^2}})_{\mu\nu} = - h^{R,{\rm log}}_{\mu\nu},\nonumber\\
 && (D^R D^R h^{R,{\rm log^2}})_{\mu\nu}  = h^{R}_{\mu\nu},\nonumber\\
 && (D^R D^R D^R h^{R,{\rm log^2}})_{\mu\nu} =  (D^R h^{R})_{\mu\nu}=0.
 \ea
The succeeding descendants of the log-square quasinormal modes at
the tricritical point 2 is simply derived by applying the
mentioned substitution, and finally yield the quasinormal
frequencies as
 \be\label{rqnms}
 \omega^n_R=-k-i4\pi T_R n, ~~n\in Z.
 \ee

\section{Log-square boundary conditions}
First of all, we review the log$^2$-boundary condition on the
AdS$_3$ spacetimes.  On the AdS$_3$ background, the log$^2$-solution
does not obey either the Brown-Henneaux boundary or the log-boundary
conditions. Hence, Liu and Sun~\cite{LS} have introduced the
log$^2$-boundary condition, \ba \label{adslogeom12} \tilde{h}^{\rm
log^2}_{\mu\nu}=
  \left(
  \begin{array}{ccc}
    \rho^2 & 1 & \rho^2e^{-2\rho}\\
    1 & 1 & e^{-2\rho} \\
    \rho^2e^{-2\rho} & e^{-2\rho} & e^{-2\rho} \\
  \end{array}
 \right)_{\mu\nu},
 \ea
which is a relaxed form obtained by replacing $ \tilde{h}_{uu}=1$
and $ \tilde{h}_{u\rho}=e^{-2\rho}$ in the Brown-Henneaux boundary
by $ \tilde{h}^{\rm log^2}_{uu}=\rho^2$ and $\tilde{ h}^{\rm
log^2}_{u\rho}=\rho^2 e^{-2\rho}$ on the AdS$_3$ spacetimes. It is
worth noting that the log$^2$-boundary condition is different from
the Brown-Henneaux boundary in the CFT.

Similarly, we conjecture that for log$^2$-quasinormal modes, its
asymptotic boundary condition should differ from those of ordinary
quasinormal modes.  In the case of quasinormal modes around the BTZ
black hole, one expects that all quasinormal modes fall off
exponentially in time $\tau$ and for large radial distance $\rho$,
together with ingoing modes at the horizon. The asymptotic behavior
for  the highest weight mode (\ref{hmassless}) for constructing the
ordinary quasinormal modes is given by~\cite{SS}
 \ba \label{ologeom12}
  h^{L}_{\mu\nu}
 &\sim&
  \left(
  \begin{array}{ccc}
    1 & 0 & e^{-2\rho}\\
    0 & 0 & 0 \\
    e^{-2\rho} & 0 & e^{-4\rho} \\
  \end{array}
 \right)_{\mu\nu}.
 \ea
On the other hand, the asymptotic behavior for  the highest weight
mode (\ref{loglogeom1}) for constructing log$^2$-quasinormal modes
takes the form
 \ba \label{loglogeom12}
  h^{L,{\rm log^2}}_{\mu\nu}
 &\sim&  \rho^2
  \left(
  \begin{array}{ccc}
    1 & 0 & e^{-2\rho}\\
    0 & 0 & 0 \\
    e^{-2\rho} & 0 & e^{-4\rho} \\
  \end{array}
 \right)_{\mu\nu}.
 \ea
Here we note that (\ref{loglogeom12}) involves  $h^{L,{\rm
log^2}}_{uu}$ which  is quadratically  growing for large  $\rho$.
This might be tempted to disqualify as a quasinormal mode because
all quasinormal modes  fall off  for large radial distance $\rho$.
Considering  Eq. (\ref{adslogeom12}), $h^{L,{\rm log^2}}_{uu} \sim
\rho^2$ in Eq. (\ref{loglogeom12}) shows a similar behavior as in
$\tilde{h}^{\rm log^2}_{uu} \sim \rho^2$ on the AdS$_3$
spacetimes. However, from the observation of the first descendent
(\ref{leftsol1}) of the log$^2$ solution, its asymptotic form is
given by
 \ba
  h^{L(1),{\rm log^2}}_{\mu\nu}
 &\sim&  \rho^2
  \left(
  \begin{array}{ccc}
  -\frac{1}{\rho} & e^{-2\rho} & e^{-2\rho}\\
    e^{-2\rho} & 0 & e^{-4\rho} \\
    e^{-2\rho} & e^{-4\rho} &  e^{-4\rho} \\
  \end{array}
 \right)_{\mu\nu},
 \ea
which contains still $h^{L(1),{\rm log^2}}_{uu}\sim \rho$-term
which gives rise to divergences at infinity.  In order to see how
this divergence is tamed, we obtain asymptotic form of the second
descendent (\ref{leftsol2}) of the log$^2$-solution
 \ba\label{2nddes}
   h^{L(2),{\rm log^2}}_{\mu\nu}
 &\sim&  \rho^2
  \left(
  \begin{array}{ccc}
   -\frac{1}{\rho} & e^{-2\rho} & e^{-2\rho}\\
   e^{-2\rho} & e^{-4\rho} & e^{-4\rho} \\
    e^{-2\rho} &  e^{-4\rho} & e^{-4\rho} \\
  \end{array}
 \right)_{\mu\nu}
 \ea
where $\rho$ still appears in the $(uu)$-element. Successively, it
is appropriate to compute the third descendants of $h^{L(3),{\rm
log^2}}_{\mu\nu}$. Its asymptotic behavior is exactly the same
with the asymptotic form (\ref{2nddes}) of the second descendent.
As a result, similar to the previous work~\cite{Kim:2012pt}, we
expect that all higher order descendants with $n>3$ behave as the
second descendent shows. This implies that one could not eliminate
the linear divergence in $h^{L(2),{\rm log^2}}_{uu}$ if one
considers the tricritical gravity, as in the log$^2$-bounary
condition on the AdS$_3$. However, this may give rise to some
difficulty in identifying the corresponding dual operator in the
CFT~\cite{Sachs}.

\section{Rank-3 LCFT}
First of all, the log gravity at the tricritical point could be
dual to a rank-3 LCFT with $c_L=0$ on the
boundary~\cite{Grumiller:2010tj,Bergshoeff:2012ev}.  The rank-3
LCFT ~\cite{Gurarie:1993xq,Flohr:2001zs} is composed of three
operators $\{ {\cal O}^{\rm L}(z),{\cal O}^{\rm log}(z),{\cal
O}^{\rm log^2}(z)\}$ which are denoted as $\{ C(z),D(z),E(z)\}$,
for simplicity. The two-point functions of these operators take
the forms
\begin{eqnarray}\label{c11}
&&<C(z) C(0)>=<C(z)D(0)>=0, \\
\label{c12} &&<C(z) E(0)>=<D(z)D(0)>=\frac{a_{\rm L}}{2z^{2h_L}},\\
\label{c13} &&<D(z) E(0)>=-\frac{a_{\rm L}\log z}{z^{2h_L}}, \\
\label{c14} &&<E(z) E(0)>=\frac{a_{\rm L} \log^2 z}{z^{2h_L}},
\end{eqnarray}
which form a rank-3 Jordan cell. Schematically, these two-point
correlation functions are represented by \be <{\cal O}^i{\cal O}^j>
\sim \left(
  \begin{array}{ccc}
    0 & 0 & {\rm CFT} \\
    0 & {\rm CFT} & {\rm L} \\
   {\rm  CFT} & {\rm L} & {\rm L}^2 \\
  \end{array}
\right),
 \ee
 where $i,j$ =$L$, log, log$^2$, CFT denotes the CFT two-point
 function (\ref{c12}), L represents (\ref{c13}), and L$^2$ denotes
 (\ref{c14}).

 At this stage, we stress that (\ref{c11})-(\ref{c14}) show a
 rank-3 zero temperature LCFT.
Even though one knows a rank-3  zero temperature LCFT, it is a
non-trivial task to construct a rank-3 finite temperature LCFT whose
zero temperature limits correspond to (\ref{c11})-(\ref{c14}). If
one knows the latter, one could read off  the log$^2$ quasinormal
frequencies from the finite temperature LCFT exactly. We note that a
rank-2 finite temperature LCFT was known in ~\cite{Myung:1999nd}.

In this section, we derive the quasinormal frequencies
$\omega^n_{L}= k- i4\pi T_L n $ (\ref{lsqnm}) of the graviton using
the finite temperature CFT. Here we wish to use the finite
temperature CFT but not the finite temperature LCFT.   In order to
derive quasinormal modes, we focus at the location of the poles in
the momentum space for the retarded two-point functions
$G^{CE}_R(\tau,\sigma)$, $G^{DD}_R(\tau,\sigma)$,
$G^{DE}_R(\tau,\sigma)$, and $G^{EE}_R(\tau,\sigma)$~\cite{Sachs}.
It is  important to recognize that as is shown in Eq. (\ref{c12}),
$G^{CE}_R(\tau,\sigma)$ and $G^{DD}_R(\tau,\sigma)$ are identical
with that of the two-point function in the finite temperature
CFT~\cite{Birmingham:2001pj}. The momentum space representation can
be read off from the commutator whose pole structure is given by
\begin{eqnarray}
&&{\cal D}^{CE}(p_+)\sim {\cal D}^{DD}(p_+) \propto
\Gamma\left(h_L+i\frac{p_+}{2\pi T_L}\right)
\Gamma\left(h_L-i\frac{p_+}{2\pi T_L}\right), \label{9}
\end{eqnarray}
where  $p_+=(\omega- k)/2$  and $T_L=1/2\pi$ for the BTZ black hole.
This function has poles in both the upper and lower half of the
$\omega$-plane. It turned out that the poles located in the lower
half-plane are the same as those  of the retarded two-point
functions $G^{CE}_R(\tau,\sigma)$ and $G^{DD}_R(\tau,\sigma)$.
Restricting the poles in Eq. (\ref{9}) to the lower half-plane, we
find one set of simple poles
\begin{eqnarray}
\omega_s&=&k- i4\pi T_L (n+h_L), \label{10}
\end{eqnarray}
with $n\in N$. This  set of poles characterizes the decay of the
perturbation on the CFT side~\cite{Birmingham:2001pj}, while
(\ref{10}) was first derived from the scalar perturbation around the
BTZ black hole~\cite{Cardoso:2001hn} and scalar wave-falloff was
discussed in AdS$_3$ spacetimes~\cite{Chan:1996yk}. At this stage,
we  would like to mention that the same quasinormal frequencies
(\ref{10}) was obtained by a slightly different operator method
based on the hidden conformal symmetry appeared in the linearized
equation ~\cite{Chen:2010ik,Chen:2010sn}, which is not an underlying
symmetry of the spacetime itself as shown in the line element
(\ref{lc}).  Sach and Solodukhin~\cite{SS} have used the latter
symmetry to construct quasinormal modes and to find quasinormal
frequencies, followed by us. Even though the symmetry group is the
same as SL(2,$R$), their origin is different.

Furthermore, $G^{DE}_R(t,\sigma)$ can be inferred by
noting~\cite{Sachs} \be <D(x) E(0)>=\frac{\partial}{\partial
h_L}<C(x) E(0)>. \ee Then, this implies  that its momentum space
representation takes the form
\begin{eqnarray}\label{ddpole}
{\cal D}^{DE}(p_+) \propto  \Gamma'\left(h_L+ip_+\right)
\Gamma\left(h_L-ip_+\right) +\Gamma\left(h_L+ip_+\right)
\Gamma'\left(h_L-ip_+\right),
\end{eqnarray}
where the prime ($'$) denotes the differentiation with respect to
$h_L$. We mention that  (\ref{ddpole}) is a relevant part for
extracting  pole structure,  but its explicit form appeared
in~\cite{Myung:1999nd}.  The poles in the lower half-plane are
relevant to  deriving quasinormal modes.  We note that ${\cal
D}^{DE}(p_+)$ has double poles, while ${\cal D}^{CE}(p_+)$ has
simple poles at the same location.  These double poles  are
responsible for explaining the linear-time dependence in $y(t,\rho)$
of the corresponding quasinormal modes (\ref{logeom1}). Restricting
the double poles in Eq. (\ref{ddpole}) to the lower half-plane, we
find one set of double  poles
\begin{eqnarray}
\omega_d&=&k- i4\pi T_L (n+h_L). \label{dpq}
\end{eqnarray}

Finally,  $G^{EE}_R(t,\sigma)$ can be inferred by noting  \be <E(x)
E(0)>=\frac{1}{2}\frac{\partial^2}{\partial h_L^2}<C(x) E(0)>, \ee
which implies  that its momentum space representation takes the form
\begin{eqnarray}\label{dddpole}
{\cal D}^{EE}(p_+) &\propto&  \Gamma''\left(h_L+ip_+\right)
\Gamma\left(h_L-ip_+\right) +2\Gamma'\left(h_L+ip_+\right)
\Gamma'\left(h_L-ip_+\right) \nonumber
\\
&+&\Gamma\left(h_L+ip_+\right) \Gamma''\left(h_L-ip_+\right).
\end{eqnarray}
 The poles in the lower half-plane are relevant to deriving
quasinormal modes. We mention that  ${\cal D}^{EE}(p_+)$ has triple
poles, while ${\cal D}^{CE}(p_+)$ has simple poles at the same
location.  These triple  poles  are responsible for the
quadratic-time dependence in $y^2(t,\rho)$ of the corresponding
quasinormal modes (\ref{loglogeom1}). Restricting the triple poles
in Eq. (\ref{dddpole}) to the lower half-plane, we find one set of
triple poles
\begin{eqnarray}
\omega_t&=&k- i4\pi T_L (n+h_L). \label{tpq}
\end{eqnarray}
All these quasinormal frequencies ($\omega_s,\omega_d,\omega_t)$ are
for the scalar with $h_L=2$. For the tensor perturbation, its
conformal weight is given by $h_L=(M-1)/2$ as (\ref{conw}) is shown.
At the tricritical point $M=1$, one has $h_L=0$ and finally,
plugging it to (\ref{tpq}) leads to (\ref{lsqnm}).

\section{Truncation of the tricritical GMG}

On the AdS$_3$ background, in order to remove the non-unitary LCFT,
one restricts the theory to the zero of Abbott-Deser-Tekin charge
($Q_L=0$). This corresponds  to truncating tricritical GMG, which
amounts to reducing  the log$^2$-boundary conditions to log-boundary
condition. After the truncation, the two-point correlation functions
take the form \be \label{tmatrix}
 <{\cal O}^i{\cal O}^j> \sim \left(
  \begin{array}{cc}
    0& 0 \\
    0 & {\rm CFT} \\
  \end{array}
\right),
 \ee
 which implies that the left-moving sector involves a non-trivial
 two-point correlator
 \be
 <D(z)D(0)>\equiv <{\cal O}^{\rm log}(z){\cal O}^{\rm log}(0)>=\frac{a_{\rm
 L}}{2z^{2h_L}}. \ee
 This is unitary and thus, the non-unitary issue is resolved  by truncating the
 tricritical point of the GMG.

What happens for the quasinormal modes when truncating the log$^2$-
quasinormal modes on the BTZ black hole? At this stage, it is not
easy to answer to this question because we did not construct  a
rank-3 finite temperature LCFT whose zero temperature limits
correspond to (\ref{c11})-(\ref{c14}). If this finite temperature
LCFT is constructed, one may apply the above truncation process to
obtain the quasinormal modes (\ref{10}) which is obtained from
simple poles existing in the finite temperature CFT. This may
correspond to CFT in (\ref{tmatrix}).

\section{Discussions}

As was mentioned in the introduction, there is a clear difference
in the AdS/LCFT correspondence between ``tricritical gravity on
the AdS$_3 \rightarrow$
 a rank-3  LCFT"
and ``tricritical gravity on the BTZ black hole $ \rightarrow$ a
rank-3 finite temperature LCFT".  We investigated the tricritical
GMG by following the latter line because the former was almost
confirmed.

We have obtained  log-square  quasinormal modes and frequencies of a
graviton  around the BTZ black hole  at the tricritical GMG  by
employing the operator method.  The log-square quasinormal
frequencies are  also obtained  by using a finite temperature CFT.
This shows the AdS/LCFT correspondence at the tricritical point
approximately.

 Even though one
knows a rank-3 zero temperature LCFT (\ref{c11})-(\ref{c14}), it
is a non-trivial task to construct a rank-3 finite temperature
LCFT whose zero temperature limits correspond to
(\ref{c11})-(\ref{c14}). If one knows the latter, one could read
off the log$^2$-quasinormal frequencies from the finite
temperature LCFT precisely.   In this case, one could apply the
truncation process to obtain the quasinormal modes (\ref{10})
which is obtained from simple poles existing in the finite
temperature CFT, together with removing (\ref{dpq}) and
(\ref{tpq}).

\section*{Acknowledgement}
Two of us (Y. S. Myung and Y.-W. Kim) were supported by the
National Research Foundation of Korea (NRF) grant funded by the
Korea government (MEST) (Grant No.2011-0027293). Y.-J. Park was
partially supported by the National Research Foundation of Korea
(NRF) Grant funded by the Korea government (MEST) through the
Center for Quantum Spacetime (CQUeST) of Sogang University with
Grant No. 2005-0049409, and was also supported by World Class
University program funded by the Ministry of Education, Science
and Technology through the National Research Foundation of Korea
(Grant No. R31-20002).

\end{document}